\documentclass{article}
\usepackage[T1]{fontenc}
\usepackage{graphics}

\makeatletter

\providecommand{\LyX}{L\kern-.1667em\lower.25em\hbox{Y}\kern-.125emX\@}

\newcommand{\lyxrightaddress}[1]{
  \par {\raggedleft \begin{tabular}{l}\ignorespaces
  #1
  \end{tabular}
  \vspace{1.4em}
  \par}
}

\usepackage[T1]{fontenc}

\makeatletter

\usepackage[T1]{fontenc}

\makeatletter

\usepackage[T1]{fontenc}

\makeatletter

\usepackage[T1]{fontenc}

\makeatletter

\usepackage[T1]{fontenc}


\makeatother
\makeatother
\makeatother
\makeatother

\begin{document}

\title{Central Baryons in Dual Models and \\
 the Possibility of a Backward Peak in Diffraction}

\author{Fritz W. Bopp \\
   \\
   Universität Siegen, Fachbereich Physik, \\
 D--57068 Siegen, Germany }

\maketitle
\vspace*{-8cm}

\lyxrightaddress{SI-00-1}

\vspace*{+7cm}

\begin{abstract}
Two distinct interactions of Pomerons should occur in dense multi-string events.
Besides the usual triple Pomeron processes transitions to membraned cylinders
can be expected to contribute in a significant way. They offer an efficient
mechanism for central baryon production and for the long range transport of
initial baryons. The slope of such an exchange should be quite low as it is
related to the Odderon known from the leading logarithmic approximation. Such
a flat trajectory has to be suppressed by small coupling constants. It is argued
that this strong suppression does not appear in diffractive events. In consequence
there should be a tiny observable backward peak in the initial baryon distribution
even in quite massive diffractive systems. 
\end{abstract}

\section{Introduction}

In heavy ion collisions the stopping of incoming baryons is stronger than expected
from a simple superposition of nucleon reactions. A number of mechanisms were
introduced in a somewhat ad hoc way to repair this shortcoming and to enhance
the slowing-down of baryons in heavy ion collisions\cite{1,2,3,4} and no real
problem seems to remain\cite{1,5}. The aim of the present work is to better
understand the basic mechanism.

In Topological models there is a definite mechanism which drastically increases
the slowing-down of baryons in heavy ion reactions. Besides the cylinder of
the conventional Pomeron scattering a membraned cylinder should appear. It could
have a significant effect on the overall event structure in dense scattering
processes. Just like the Pomeron it can be understood as a soft extrapolation
of a known hard QCD exchange. If observed this second example of a connection
between a soft and a hard object would be of considerable theoretical interest.

Unfortunately there is a large quantitative uncertainty. The relevant slope
parameter is not sufficiently constrained by existing baryon exchange data to
allow definite conclusions about the expected weight of this new contribution.
However, there is a very specific effect in diffractive final states and possibly
in electro-production. The existence of such a contribution can therefore be
tested and its weight is accessible to direct experimental observation.

After a short excursion to hadron hadron scattering we will consider heavy ion
reactions and discuss the proposed mechanism of baryon transport in this context.
We will then turn to general consequences including the clear cut backward scattering
peak in diffractive events.

\section{Baryon transfer in particle scattering }

\paragraph{Available experimental data}

To observe the slowing down of the incoming baryon charge to central or opposite
rapidities, one needs to somehow identify the incoming contribution to the baryon
spectra. It is usually assumed that the produced baryons and anti-baryons have
identical distributions. A simple subtraction then provides the desired initial
baryon distribution. This trick does not work for proton antiproton scattering,
where the sea-baryon contribution cannot be determined from data without model
assumptions. Unfortunately this precludes the use of post ISR data excepting
HERA and diffractive systems with sufficiently large Pomeron-proton sub-energies. 

Available are spectra from meson-baryon processes (compiled in \cite{6}) and
for a suitable combination of proton-proton and proton-antiproton processes
(compiled in \cite{7}). The incoming proton spectrum at fixed \( p_{\bot } \)
(plotted in \cite{8} or \cite{9}) are consistent with a slope of \( \alpha _{Transfer}-\alpha _{Pomeron}=-1 \)
with large error. As the central data points are somewhat on the high side there
is a hint of an eventual turnover to a flatter slope. The \( p_{\bot } \)-integrated
ratio of the incoming baryon and the sea-baryon distribution was given in \cite{10,11}
to be:

\[
A_{ISR}=\frac{\rho _{initial\, baryon\, charge}(y)}{\rho _{sea\, baryon\, charge}(y)}=0.39\pm 0.05,\: 0.33\pm 0.05,\: 0.23\pm 0.05\]
 for \( y=-0.4 \), \( y=0 \) and \( y=+0.4 \)\footnote{
As the data have anyhow a large error the numbers are just taken from the figure
29\cite{10}. The subtraction assumed that the sea-antiproton distributions
in proton-antiproton and proton-proton scattering are equal.
}. The central derivative of \( A_{ISR} \) obtains no contribution from the
(symmetric) sea-baryon distribution. Assuming the usual exponential distribution

\begin{equation}
A\propto \exp [(\alpha _{Transfer}-\alpha _{Pomeron})y]
\end{equation}
 the quantity 
\begin{equation}
\frac{d/dy\, A_{ISR}}{A_{ISR}}|_{y=0}=\alpha _{Transfer}-\alpha _{Pomeron}=-0.49_{-0.37}^{+0.42}
\end{equation}
 just yields the slope\footnote{
A similar value was obtained in\cite{2} in a fit which required more assumptions.
} . While not in absolute contradiction with a slope around one the indicated
value is again considerably less.

New preliminary data come from the H1 experiment at HERA\cite{12}. They observed
the initial baryons asymmetry at laboratory rapidities

\begin{equation}
A_{H1}=\frac{\rho _{initial\, baryon\, charge}(y)}{\rho _{sea\, baryon\, charge}(y)}=0.08\pm 0.01\pm 0.025
\end{equation}
 A simple extrapolation of the ISR value to the larger rapidity difference\footnote{
It assumes that the sea stays constant. With an increasing sea baryon spectrum
the conclusion about the flattening slope would be slightly stronger.
} would have \char`\"{}predicted\char`\"{}: 

\begin{equation}
A_{H1}=0.061_{-0.046}^{+0.243}
\end{equation}
 Hence the H1 values lie within the expected range . The required rather flat
value of the slope is
\[
\alpha _{Transfer}-\alpha _{Pomeron}=-0.4\pm 0.2\]

The baryon stopping seen in the spectra is related by the Mueller-Kancheli relation
to the annihilation cross section. With certain assumptions similar conclusions
yielding a steep slope and a possible flattening at higher energies can be drawn
from this data \footnote{
Data on identified annihilation are available only at energies below \( 20 \)
GeV. The data fall of like a power of roughly \( a_{Transfer}-1=-1 \). In this
range the difference in the proton-proton and proton-antiproton cross sections
is saturated by annihilation. At high energies this difference in the cross
sections turns to a flatter slope of roughly \( a_{Transfer}-1=-0.5 \). The
data\cite{13,14} have smaller errors and the indication of a knee is stronger
than in the inclusive ISR distribution. However the interpretation as annihilation
process is not clear as mesonic trajectories (\( \omega +f_{0} \) ) also contribute
to \( p\bar{p} \) scattering only. 
}.

\paragraph{Dual Topological picture }

There are several Regge-pole contributions for the slowing down of baryons in
hadron hadron scattering. The basic philosophy of the Dual Topological models\cite{15}
in the classification of such exchanges\cite{16} involves ``materializing''
or ``suppressed'' strings. ``Materializing'' means that the initial color
fields are neutralized by a chain of hadronizing \( qq \) pairs, ``suppressed''
means hadron-less neutralization by an exchange of a single quark. It is analogous
to the Pomeron and the Reggeon exchange where in addition to a two chain Pomeron
a one-chain Reggeon has to be considered. Phenomenologically contributions with
various suppressed strings have to be considered as independent and additive.
For each suppressed string an extra factor \( (\sqrt{1/M_{string}}) \) appears
and restricts the suppressed contribution to low energies.

For a nuclear exchange one starts with a completely suppressed exchange, i.
e. with the square of the quasi-elastic nucleon exchange amplitude
\begin{equation}
\alpha _{junction}^{III}-1=2(\alpha _{Nucleon}-1)=-2
\end{equation}
 known from elastic backward scattering. Each of the three exchanged valence
quarks can now be replaced by a ``materializing'' string. Corresponding to
three, two, one or zero strings there are four contributions with trajectories
spaced by one half. At considered energies the first two of these ``baryonium''
trajectories with two and three hadronizing strings
\begin{equation}
\alpha _{junction}^{0}-1=-0.5,\: \alpha _{junction}^{I}-1=-1.0
\end{equation}
 will be relevant. They could be responsible for the initially steep (\( -1.0 \))
and then possible flattening (\( -0.5 \)) slope observed in the data discussed
above.

The value of the final slope is rather uncertain. Even a value of \( \alpha _{junction}^{I}-1=0 \)
was proposed in the literature\cite{17,18}. The correspondence to the Odderon
gives some support to such a value. Another uncertainty comes from the \( \omega  \)-trajectory.
The baryonium exchange has the same quantum numbers as the \( \omega  \)-meson
Reggeon. A simple estimate of an additional \( \omega  \)-contribution leads
to a too large non-annihilation contribution\cite{18}. One solution is to identify
the initial trajectory as a mixture of both contributions was developed in literature\cite{19,20}.
In this way it is possible to identify the turnover in the slope with onset
of the sea baryon antibaryon production observed in the inclusive spectrum \cite{9}.
The predicted strong correlation between baryon stopping and sea-baryon antibaryon
density is not found in the data\cite{10} and we therefore assume here that
the mixing if existing is very weak. An alternative solution to the \( \omega  \)-problem
will be given below.

\paragraph{Implementation in Dual Parton model based Monte Carlo codes}

The Dual Parton model was developed for high energies and it has in its present
implementation no mesonic Regge-pole exchanges appear in the iterations which
determine the cross sections.\footnote{
To be precise, some Monte Carlo implementations (like DPMJET\cite{5} or PHOJET\cite{21})
include simple Regge exchanges to stay applicable at lower energies. Also if
individual chains happen to come out too light to produce partons their parent
quarks are annihilated to mimic a Regge-pol exchange as far as the final state
is concerned. The neglect of the (flavor moving) Regge contribution is questionable
in heavy ion scattering when multiple scattering processes are common and sub-energies
of the involved constituents are often quite low. 
}. Considering just the interplay of local baryonium exchanges within a global
Pomeron exchange is straight forward. The factorization among strings allows
to ignore the quark string which is common to both trajectories. The transition
of the remaining diquark string (baryonium remnant) into an antiquark string
(Pomeron remnant) can be implemented in a usual fragmentation scheme by a suitable
choice of the splitting function for diquark-diquark, quark-diquark, diquark-quark
and quark-quark transitions. It was implemented in most string models e.g. \cite{22,23,24}
and it is part of the JETSET program (as diquarks or as pop-corn mechanism\cite{25}).
Without relying on string factorization leading- and sea-baryon exchanges are
also implemented in HIJING/\( B\bar{B} \). \cite{26}.

\section{Baryon enhancement \\
 in dense heavy ion scattering}

\paragraph{Concepts for slowing-down initial baryons }

There are a few completely conventional mechanisms of baryon transfer and central
baryon production for multiple scattering processes in string models\footnote{
Besides the purely kinematic \char`\"{}attenuation\char`\"{} effect\cite{27}
in multiple scattering events, the initial baryon can be slowed down if a secondary
Pomeron exchange picks up a valence quark and leave the diquark with one (typically
slower) sea quark\cite{28}. An obvious mechanism of central baryon production
involves sea diquark-antidiquark pairs. Except for a limited number of valence
partons, strings have to connect to sea partons which can also be diquark-antidiquark
pairs\cite{29}. It is known from the analysis of the transverse momenta, i.e.
of \( <p_{\bot }>_{n_{\{charged\}}} \) \cite{30}, that the partons of the
string ends come from a harder (naturally more \( SU(3) \) -symmetric) initial
phase. Multiple additional sea quarks are therefore helpful in the understanding
of the strangeness enhancement. 
}. They were investigated nu\-mer\-i\-cal\-ly\cite{5}, they are helpful in some
regions but not enough to explain the large stopping in heavy ion scattering.

To understand the data it seems necessary to include interplay of string if
they get sufficiently dense in transverse space. It was proposed that there
are new special strings\cite{31,32}. In contrast, we shall maintain here the
general factorization hypothesis between initial scattering and the final hadronization
within standard strings. We will just consider a more complex string structure. 

The usual Pomeron exchange in the Dual Parton model leaves a quark and a diquark
for the string ends. Diquarks are no special entities and multiple scattering
processes have no reason not to split them in a suitable conventional two Pomeron
interaction. It is natural to expect that diquark break-ups considerably slow
down the baryons evolving. The probability for such an essentially un-absorbed\cite{33}
process is\cite{34}:
\begin{equation}
{[break\: up{]}/{[}no\: break\: up{]}\propto {[}cut\: Pomeron\: number{]}-1}
\end{equation}
 As required by the experimentally observed slowdown this is a drastic effect
for heavy ion scattering while for hadron-hadron scattering multiple scattering
are sufficiently rare (especially at energies studied in detail) to preserve
the known hadron-hadron phenomenology. How such processes are affected was considered
numerically in \cite{5} and no manifestly disturbing effects were found.

We emphasize here that the behavior of the baryon quantum number slowed down
by such a break-up is not trivial. In topological models the baryon contains
Y-shaped color electric fluxes. Two Pomerons intercepting two different branches
will leave two ``free'' valence quarks and a valence quark connected with
the vortex line with the velocity of initial Baryon to form the end of the strings.
Nothing is a priory known about the energy distribution of such quarks with
vortex lines in the structure function. A simple identification with the attached
quark distributions\cite{36} somewhat in the spirit of color evaporation models
has no basis in a dual framework.

\paragraph{Special baryon transfers in the Topological model}

For a more detailed description of the slowing down we turn to the Dual Topological
model\cite{16} introduced in section 2. A discussion of baryon transfers in
such a framework was recently given by Kharzeev\cite{37}. We will here emphasize
topological aspects.

In topological models a Pomeron exchange corresponds to a cylinder of in a certain
way arranged gluon fields connecting the two scattering hadrons. If one considers
an arbitrary plane intersecting this exchange (say at a fixed exchange-channel
time) the intersection of the cylinder is topologically equivalent to a circle.
More specifically in topological models amplitudes with clockwise respectively
anticlockwise orientation have to be added or subtracted depending on the charge
parity. The cylinders or the circles therefore come with two orientations. This
distinction is usually not very important as it is always possible to attach
hadrons in a matching way; except for C-parity conservation no special restrictions
result. 

Pomerons have a transverse extent and if they get close in transverse space
they should interact. Hadronic interaction is sufficiently strong to be largely
determined by geometry. It is therefore reasonable to expect that the coupling
does not strongly depend on the orientation as long as there is no mechanism
of suppression. 

The two distinct configurations lead to different interactions. Two Pomerons
with the same orientation can if they touch (starting locally at one point in
the exchange-channel time) shorten their circumference and form a single circle.
This then corresponds to the usual triple Pomeron coupling experimentally well
known from diffractive processes. 

For two Pomerons with opposite orientation the situation is more complicated.
Like for soap bubbles the two surfaces which get in contact can merge and form
a single membrane. The joining inverts the orientation of the membrane. On the
intersecting plane one now obtains -- instead of the single circle -- three
lines originating in a vortex point and ending in an anti-vortex point as shown
in figure 1 . 
\begin{figure}
{\par\centering \resizebox*{0.5\textwidth}{!}{\includegraphics{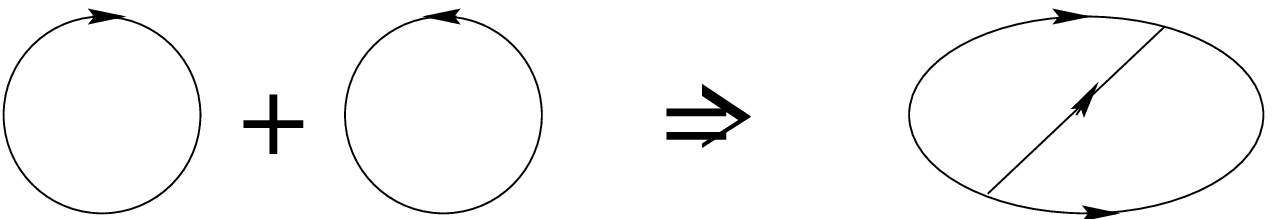}} \par}

\caption{The joining of two cylinders with opposite orientation}
\end{figure}
Lacking a topological name for the three dimensional object the term membraned
cylinder will be used in the following.

How do this membraned cylinder contribute to particle production? Similar to
the triple Pomeron case there are three different ways to cut through a membraned
cylinder. The cut which also intersects the membrane has vortex lines on both
sides. They present a topological description of the baryon transfers considered
in this paper. Cuts which intersect only two sheets contribute as an absorptive
reduction of the two string contribution. Their negative contributions make
the understanding of total cross sections not straightforward. It is possible
that the membraned cylinder exchange has a vanishing or negative imaginary part.
The \( \omega  \)-exchange could indeed dominate the difference in the cross
section while the contribution of the three string cut of the membraned cylinder
could be more or less compensated by its negative two string cut. In the final
states the annihilation process can be observed while the one string \( \omega  \)-exchange
contribution and the two string membraned cylinder reduction is hidden resp.
contained in the usual two string contribution.

\paragraph{The identification with the Odderon}

Even though QCD cannot presently be used to calculate soft processes the typical
absence of a abrupt changes in experimental distributions indicates that there
is no discontinuous transition between soft and hard reactions both formulated
on a partonic level. This provides the hope that hard processes can be used
as a guide and that soft processes can be parameterized as an extrapolation
of calculable hard processes. 

The well known example is the connection between soft and hard Pomerons. To
identify the hard partner of the soft Pomeron we first observe that the simplest
representation of a Pomeron in PQCD involves the exchange of two gluons. As
spin one particles exchanged gluons introduce no energy dependence and two gluons
can form color singlets with the required positive charge parity. Following
this concept it can be shown\cite{38} that a generalization of such an exchange
gives in a rather well defined approximation the dominant contribution at very
high energies. It involves a ladder of Reggeized gluons and is called the ``hard''
or BFKL Pomeron. Topologically in a leading \( 1/N \) -expansion gluons can
be represented by pairs of color lines drawn without crossing on a geometrical
structure representing the considered contribution to the amplitude. In this
expansion the leading structure of a BFKL Pomeron corresponds to a cylinder
with the two basic gluons exchanged on opposites sides parallel to the axis.
As they are in a color singlet state their matching color lines can be connected
in front of the cylinder as shown in Fig.2a . Analogously the outer lines can
be connected on the back of the cylinder. 

Going back to the soft regime the basic assumption in topological models is
that the \( 1/N \) -expansion stays valid \footnote{
The \( 1/N \) expansion is in principle destroyed in the soft limit by huge
combinatorial factors if the number of considered gluons increases. The hope
is that preconfinement effects and a typical correlation of spatial coordinates
and momenta create a situation where the (stray) long distance gluon exchanges
which is responsible for the disturbing combinatorial factors cancel as color
and anti-color lines are too closely neighbored when seen from a distance. 
} and that the soft Pomeron therefore maintains its cylindrical structure. If
cut, soft and hard Pomerons therefore lead to similar two string final states.
As difference it remains that the slope of the soft Pomeron is just shifted
downward roughly a third of a unit.

\begin{figure}
{\par\centering \resizebox*{0.5\textwidth}{!}{\includegraphics{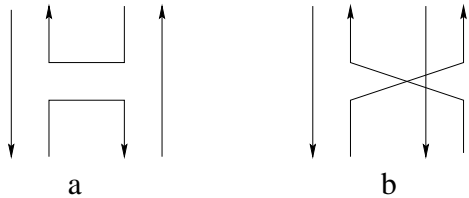}} \par}

\caption{The color lines of a gluon link between two exchanged gluons}
\end{figure}

Can one find a similar connection for the membraned cylinder? The simplest representation
spanning such a structure involves at least three gluons, one on each sheet
again parallel to the axis. Any gluon connecting these exchanges has then to
pass through a vortex line. In the \( 1/N \) expansion this means that the
color lines have to cross like in Fig. 2b\footnote{
In spite of the crossing of color lines it is a leading contribution in the
\( 1/N \) expansion. The introduction of baryons in this expansion has to be
done with care\cite{16}. For a given \( N \) a membraned cylinder would actually
require a structure with \( N-2 \) membranes with \( N \) gluons so that suppression
( \( 1/N \) ) associated with the crossed exchange would be compensated by
the \( N-1 \) connection choices.
}. A color singlet of three gluons can have the quantum numbers of a Pomeron
or an Odderon\cite{39}. There is a simple topological property of the Odderon.
A single gluon connection of type Fig.2a would project the color structure of
the pair to that of a single gluon (or singlet) and the exchange would have
to correspond to a Pomeron-like effective two gluon contribution. The Odderon
will therefore have to involve only crossed connections shown in Fig.2b. Hence
it has the topology of the membraned cylinder.

The identification with the Odderon fixes the C-parity of the membraned cylinder
and a twisted membraned cylinder will have to contribute to \( pp \) scattering
at least asymptotically equally and with opposite sign as in \( p\overline{p} \) scattering.
For the twisted exchange only one type of cut exists, it will always involve
a two step transition from a vortex-antivortex piece to a two string piece with
both vortices on one side and to an antivortex-vortex piece. Depending on its
sign it will absorb or add to the contribution in which the two string part
is replaced by a usual Pomeron cut. In diffraction the situation is more complicated.
As the imaginary (and real) part of the Pomeron-Pomeron-Odderon contribution
has to vanish, the sum over all contributions will cancel. As some cuts have
a negative sign no restriction on individual terms in the sum result. 

In the same QCD approximation as the ``hard'' Pomeron the properties of a ``hard''
or BKP Odderon\cite{40} were calculated. The consensus is that the corrections
to the initial gluon intercept are smaller than for the Pomeron and the intercept
is rather firmly predicted to be close to \( 0.96 \)\cite{41}. There is a
mismatch between this hard Odderon intercept and the (with large error) experimentally
observed soft value again by about a third of a unit.

\section{Experimental consequences \\
 of membraned cylinder exchanges.}

\paragraph{Odderon in heavy ion scattering }

In heavy ion scattering where the Pomerons are dense in transverse space they
can join and form a Pomeron or a membraned cylinder. This helps with the problem
of unphysical high string densities. The individual strings pairs are no longer
independent but the general picture of particle production in separate universal
strings survives. There should be a considerable probability of membraned cylinder
exchanges growing proportional to the density:
\begin{equation}
\frac{{[}number\: of\: membraned\: cylinders{]}}{{[}Pomeron\: number{]}}\propto \frac{{[}Pomeron\: number{]}{[}Pomeron\: radius{]}^{2}}{{[}nucleus\: radius{]}^{2}}
\end{equation}
 The transition from a Pomeron pair to the centrally cut membraned-cylinder
requires a baryon antibaryon pair production. Between a proton and a Pomeron
the cut membraned-cylinder is a very efficient mechanism of baryon stopping. 

As the slope is not well determined it is hard to obtain really reliable quantitative
statements which can be tested with convincing results in heavy ion scattering.
There is however a very specific qualitative prediction which should be testable.

\paragraph{The backward peak in diffraction and possibly in electro production}

We consider a diffractive system whose mass exceeds ISR energies. Usually the
diffractively produced particles will originate in two strings of a cut Pomeron
and the baryon charge will stay on the side of the initial proton. As usual
there might be some migration to the center with a slope eventually corresponding
to the difference of the Odderon and the Pomeron trajectory.

To accept the high soft Odderon slope suggested by the data on baryon transfers
it is necessary to require a clear suppression from the coupling constants.
It is quite natural to assume that the transition between a baryon exchange
(cut membraned cylinder) and non baryon exchange (cut Pomeron) has no large
overlap and a relatively small coupling. No such suppression is expected at
a two Pomeron vertex. This has a direct consequence in rapidity space. At a
certain distance it should be more favorable for an Odderon to span the total
diffractive region and to utilize in this way the more favorable coupling to
the two Pomerons. In consequence the initial baryon will end up exactly at the
end of the string. In the usual presentation of rapidity plots this might be
smeared out if different masses of diffractive system are included. This problem
can be solved if one plots the rapidity distribution of initial baryons in relation
to the inner end of the diffractive region, i.e. as function of 

\[
y_{\{Pomeron\}}=y_{\{CMS\}}-\ln \frac{m\sqrt{{s}}}{M(diffr.)}\]

The expectation is that a small backward peak should then be visible. To substantiate
this we show in Figure 3 the result of a calculation with the PHOJET Monte Carlo
code\cite{21} with standard parameters. To select diffractive events a lower
cutoff of \( x_{F}=0.95 \) was used.
\begin{figure}
{\par\centering \includegraphics{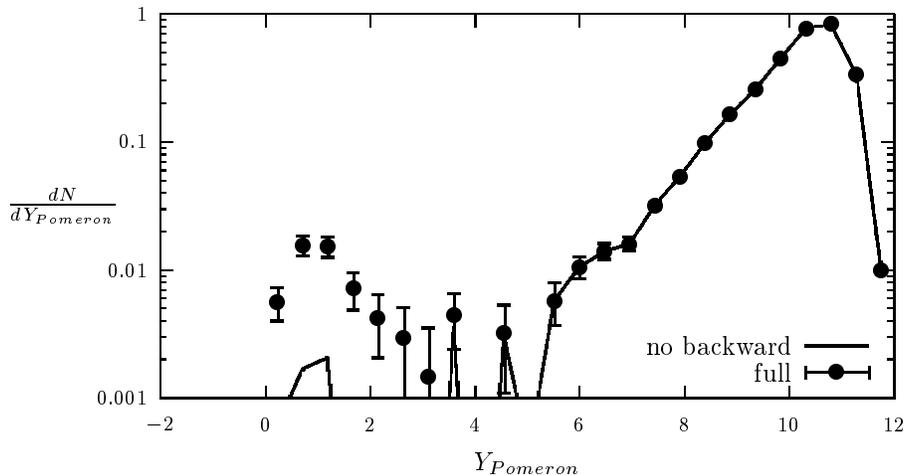} \par}

\caption{The incoming proton spectrum for diffractive events with a mass of \protect\( 300\protect \)
GeV generated with a modified PHOJET generator for \protect\( pp\protect \)-scattering
of \protect\( 1.8\protect \) TeV.}
\end{figure}
PHOJET contains diquark exchanges and yields reasonable baryon spectra in the
forward region. To obtain the postulated backward peak we just mix a suitable
sample of such events. These special events are obtained by suitably inverting
the rapidity distribution of usual events. For this inverted contribution the
diquark exchanges were disabled to ensure that the backward baryon ends up on
the last rank. 

For a diffractive system with a mass of \( 300 \) GeV the suppression from
the slope is of the order of a factor of \( 10 \). To account for the unknown
ratio of coupling constants and for uncertainties connected with the somewhat
simple implementation we add a factor \( 6 \) and took \( 30 \) million normal
events and \( 0.5 \) million inverted ones. It is important to stress that
this is not an absolute prediction but an illustration of the reasonably expected
effect. No effort was made to explicitely include the turnover in the baryon
spectra discussed in section 2. 

It is clear that suitable diffractive events should be available at the TEVATRON
and that such data could be decisive. Similar measurements might also be possible
at HERA. It is likely in the scattering of a virtual photon on a proton that
the photon does not prefer a fixed topology and the coupling to an Odderon is
also not disfavored. In this case the same backward peak might be observable
in non diffractive \( ep \) data.

\section{Conclusion}

The present paper wants to encourage the measurement of the initial baryon distribution
in high mass diffractive systems. The known initial baryon distribution suggests
that there is a contribution to the exchange which is most strongly suppressed
by its coupling and not by its slope. It is argued that this should not be the
case in diffractive events and in consequence there should be an observable
tiny backward peak in the initial baryon distribution in the diffractive system.
The prediction is important as it has manifest consequences for heavy ion processes,
where it would be a strong mechanism for central baryon production and for the
transport of initial baryons to the central and opposite region. It might also
clarify the role of the Odderon.

\section*{Acknowledgments}

I thank Lech Szymanowski and L. Lukaszuk for discussion regarding the Odderon.
I am indebted to J. Ranft and H. Anlauf for reading the manuscript and many
useful comments. I also acknowledge his help in running the PHOJET code to obtain
Figure 3. The work is based on work done with Patrick Aurenche.

\end{document}